\documentclass[twocolumn]{aastex631}

\shorttitle{M67 Blue Stragglers with High-Resolution Infrared Spectroscopy}
\shortauthors{Brady et al.}
\graphicspath{{./}{figures/}}

\usepackage{natbib}
\usepackage{xcolor}
\usepackage{amsmath}

\def\teff{\mbox{T$_{\rm eff}$}}
\def\logg{\mbox{log~{\it g}}}

\DeclareUnicodeCharacter{2212}{-}

\begin{document}

\title{M67 Blue Stragglers with High-resolution Infrared Spectroscopy}

\author[0000-0003-3271-9434]{K. E. Brady}
\affiliation{Department of Astronomy, Indiana University, Bloomington, IN}

\author[0000-0002-3456-5929]{C. Sneden}
\affiliation{Department of Astronomy and McDonald Observatory, University of Texas, Austin, TX}

\author[0000-0002-3007-206X]{C. A. Pilachowski}
\affiliation{Department of Astronomy, Indiana University, Bloomington, IN}

\author[0000-0002-2516-1949]{Melike Af\c{s}ar}
\affiliation{Department of Astronomy and Space Sciences, Ege University, \.{I}zmir, Turkey}
\affiliation{Department of Astronomy and McDonald Observatory, University of Texas, Austin, TX}

\author[0000-0001-7875-6391]{G. N. Mace}
\affiliation{Department of Astronomy and McDonald Observatory, University of Texas, Austin, TX}

\author[0000-0003-3577-3540]{D. T. Jaffe}
\affiliation{Department of Astronomy and McDonald Observatory, University of Texas, Austin, TX}

\author[0000-0002-8443-0723]{N. M. Gosnell}
\affiliation{Department of Physics, Colorado College, Colorado Springs, CO}

\author{R. Seifert}
\affiliation{Department of Astronomy and McDonald Observatory, University of Texas, Austin, TX}



\begin{abstract}
We report on the first detailed infrared chemical analysis of five binary members (S277, S997, S975, S1031, and S1195) in the open cluster M67 (NGC 2682). These stars are located outside (bluer and/or brighter than) the main sequence turnoff region in M67. High-resolution (R $\sim$ 45,000) near-infrared spectra were obtained with the Immersion GRating INfrared Spectrograph (IGRINS) at the McDonald Observatory 2.7m Harlan J. Smith Telescope, providing full spectral coverage of the $H$ and $K$ bands. Abundances of C, Na, Mg, Al, Si, S, Ca, Fe, and Ni are measured using neutral atomic absorption lines. We detect $v$ sin $i$ $\geq$ 25 km s$^{-1}$ in three of our program stars: S1031, S975, and S1195. We find our derived abundances to be in good agreement with turnoff star abundances, similar to published analyses of blue straggler stars in M67 from optical spectra. Detection of a carbon enhancement or depletion resulting from mass transfer is difficult due to the uncertainties in the carbon abundance and the relatively modest changes that may occur through red giant and asymptotic giant branch evolution. 
\end{abstract}



\section{INTRODUCTION}\label{intro}
Blue straggler stars (BSSs) are classified as stars more luminous and/or bluer than the main sequence (MS) cluster turnoff. These enigmatic stars appear as an extension of the main sequence on an optical color-magnitude diagram (CMD), seemingly lagging behind the standard stellar evolutionary track. Since their first detection in the globular cluster M3 \citep{sandage53}, BSSs have been identified in other environments including open clusters  \citep[e.g.,][]{ milone94, ahumada07}, the Galactic field \citep[e.g.,][]{preston00}, and dwarf spheroidal galaxies \citep[e.g.,][]{momany07}. Now BSSs are accepted as components of all old stellar populations.

A variety of scenarios have been invoked to account for the origin of these late bloomers. Current theories for BSS formation focus on binary and triple star processes and include: \textit{(i)} mass transfer through Roche-lobe overflow in binary systems \citep{mcrea64}; \textit{(ii)} stellar collisions during dynamical encounters \citep{hills76, leigh11, geller13}; and \textit{(iii)} mergers of inner binaries in hierarchical triples driven by the Kozai mechanism \citep{perets09, naoz14}. Each formation mechanism is most likely responsible for a portion of the BSS population, though the distribution remains unknown \citep{stryker93,leonard96}. Furthermore, the dominant formation pathway may depend on the environment. In less dense environments, such as open clusters and in the field, collisions of single stars are less likely than in high density environments \citep{stryker93}.

High-resolution infrared (IR) spectroscopy of the anomalous stars in clusters may provide additional evidence to constrain their formation, from the direct detection of cool, low-mass companions or by the derivation of evolutionary-sensitive elements of the CNO group. A particular advantage of IR spectra of binary stars is that the spectra are less likely to be altered by flux from hot companions. 

Analyzing the surface compositions of BSSs may give insight on how they were formed. If formation occurs through mass transfer, the material gained may come from the stellar interior of the donor where abundances have been modified by CNO-cycle fusion. Depending on the type of mass transfer, normal carbon, nitrogen, and oxygen abundances or modest depletions or enhancements of these elements can be expected \citep{sarna96}. When mass transfer occurs while the donor is on the main sequence it is classified as Case A, after H exhaustion but before core ignition of helium as Case B, and after the ignition of helium as Case C \citep{lauterborn70,kippenhahn67}. Conversely, if a blue straggler forms through a dynamical collision of two stars which results in a merger, minimal mixing occurs \citep{lombardi95}. \citet{lombardi02} indicates that the surface material from a merger comes from the more massive star.

\defcitealias{geller15}{G15}

The old ($\sim$4 Gyr), solar metallicity open cluster M67 (NGC 2682) is close in proximity (0.85$\pm$0.04 Kpc \citep{angelo19}) and contains a significant population of blue straggler stars \citep[][hereafter G15]{geller15}. Most BSSs are found to be binaries; \citetalias{geller15} detected a binary frequency of 79\% $\pm$ 24\% among the M67 blue stragglers. This cluster's blue straggler candidates, such as the 24 stars from \citet{deng99}, have been studied extensively through photometric \citep[e.g.,][]{gilliland91, stassun02, sandquist03, pandey21}, spectroscopic \citep[e.g.,][]{mathys91, latham96, shetrone00, liu08, motta18}, and theoretical \citep[e.g.,][]{hurley01, hurley05} investigations. 

Few of M67’s blue stragglers, however, have been analyzed chemically. Previous M67 BSS abundance analyses include \citet{mathys91}, \citet{shetrone00}, and \citet{motta18}, in which 11 BSSs in total were analyzed. These chemical analyses all found the blue straggler surface abundances to be consistent with those of turnoff stars, which is indicative of a collisional formation mechanism \citep{lombardi95}. 

We have gathered high-resolution near-IR spectra of five binary members of M67 located blueward of and/or more luminous than the MS turnoff, with \citet{sanders77} identifications of S277, S997, S1031, S975, and S1195. Section \S\ref{obsred} describes the target selection, a literature review of our targets, observations, and data reduction. Section \S\ref{modabund} details the line list construction, atmospheric parameter derivation, abundance analysis, NLTE corrections, and results. In Section \S\ref{discuss}, we discuss the rotational velocity of our program stars and put into context what our chemical composition results could mean for the stars' formation. In Section \S\ref{conclusions}, we summarize the main conclusions from our study. 

\section{OBSERVATIONS AND DATA REDUCTION}\label{obsred}

\subsection{Star Selection and Photometry}

The program stars were selected from \citetalias{geller15}'s M67 cluster membership study. \citetalias{geller15} presented radial-velocity (RV) measurements for 1278 candidate members of M67 and computed membership probabilities for stars with $\geq$ 3 RV observations. They sorted the candidate members into eight qualitative membership classes: single members/non-members, binary members/non-members, binary likely members, binaries with unknown RV membership, binary likely non-members, and unknown RV membership (see Table 3 in \citetalias{geller15} for the selection criteria for each class). Binaries were identified by having an $e/i$ statistic $\geq$ 3, or having a binary orbital solution. The $e/i$ statistic is the ratio of the standard deviations ($e$) to the expected precision ($i$) of the RVs for a given star. Stars with an $e/i$ $<$ 3 and without a derived binary orbital solution were classified as single stars, though \citetalias{geller15} note that a number of these stars may be in long-period binaries. 

\begin{figure}[t]
\epsscale{1.178}
\plotone{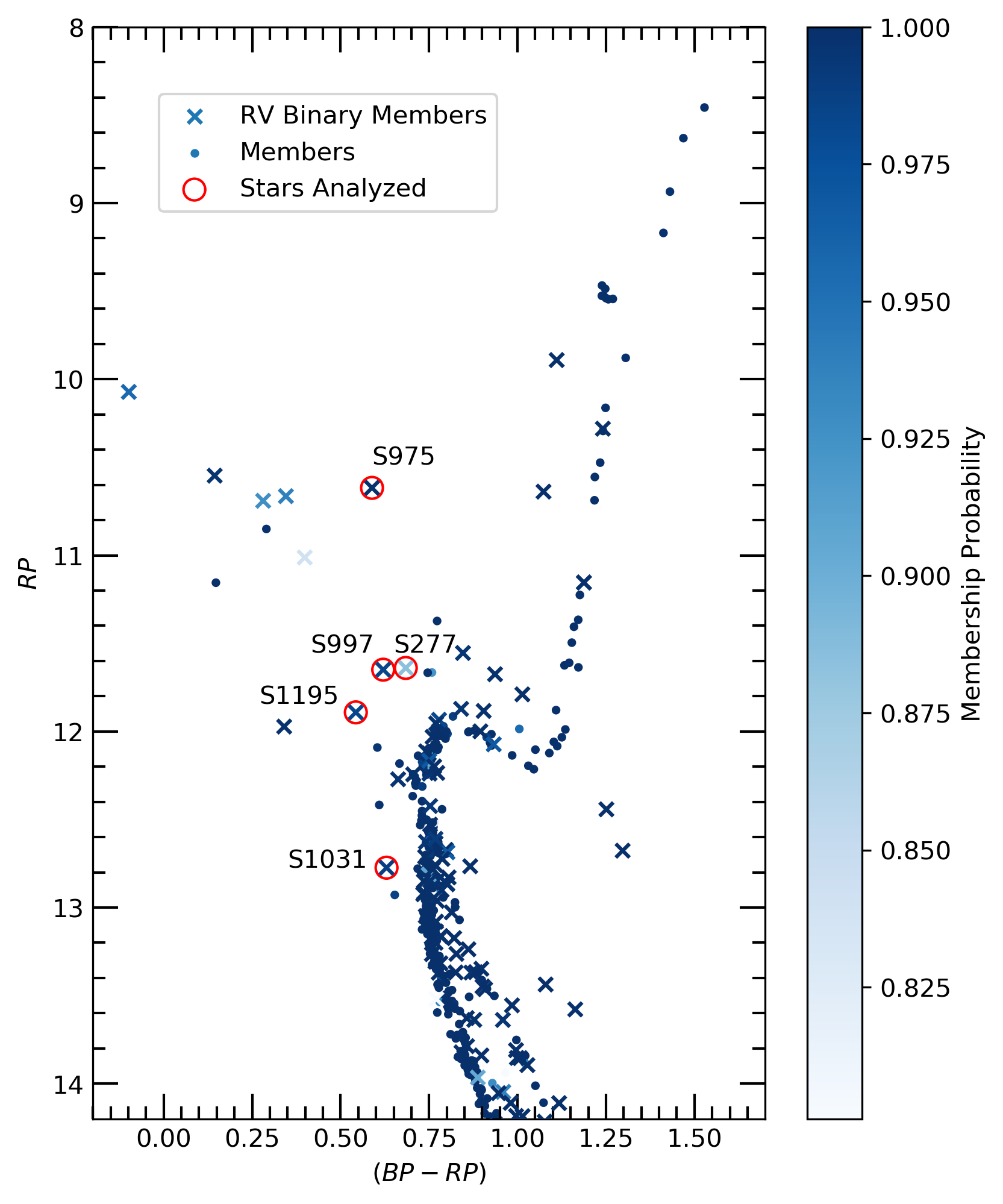}
\caption{Optical CMD of M67 with designated members by \citetalias{geller15} and photometry from Gaia DR2. RF cluster membership probabilities from \citet{gao18} are indicated with the symbol shade of blue as defined by the color palate, with the exception of S277, S997, and S975, which are RV membership probabilities from \citetalias{geller15}. The program stars are circled.\label{figure:1}}
\end{figure}

Figure \ref{figure:1} shows a CMD with the designated single members, binary members, and binary likely members from \citetalias{geller15} with Gaia Data Release 2 (DR2) photometry \citep{gaia18}. Membership probabilities are from \citet{gao18}, with the exception of S277, S997, and S975, where the membership probability is from \citetalias{geller15}. To calculate membership probabilities, \citet{gao18} used a machine-learning method, random forest (RF), based on astrometry and photometry taken from Gaia DR2. Table \ref{table:1} displays the program stars, spectral types, equatorial coordinates, and basic photometry from \citetalias{geller15}. 

\begin{figure}[ht]
\epsscale{1.121}
\plotone{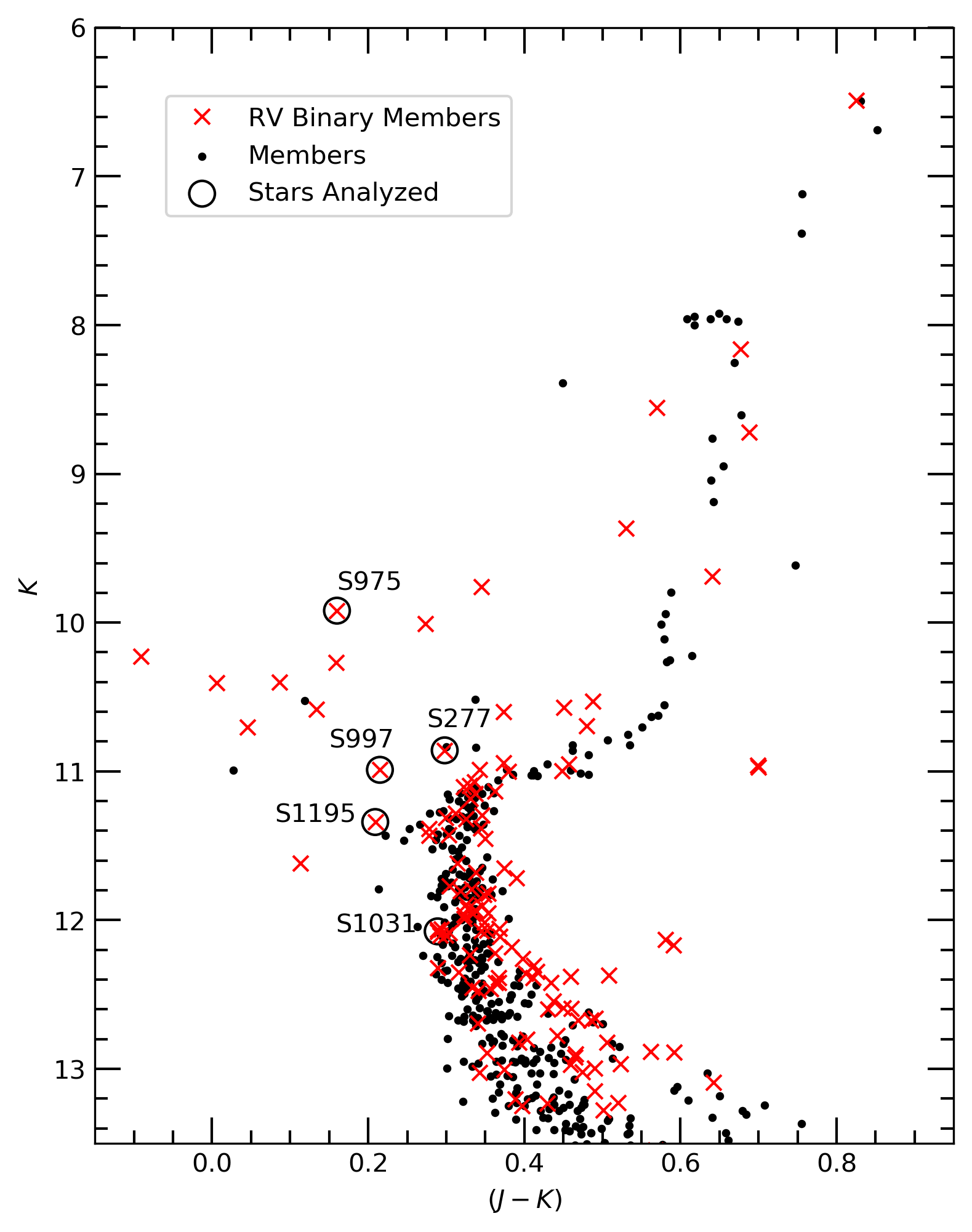}
\caption{Infrared CMD of M67 with designated members by \citetalias{geller15} and photometry from 2MASS \citep{cutri03}. The program stars are circled. \label{figure:2}}
\end{figure}

IR photometry can provide further information on BSS binaries, as the type of companion to a blue straggler influences the location of the system on CMDs of different wavelengths. Hot companions may distort colors on an optical CMD, but will have less effect on an infrared one. Figure \ref{figure:2} shows a $K$ vs. $J-K$ CMD of M67, using the infrared photometry from the 2MASS All Sky Catalog of point sources \citep{cutri03}. All of our program stars remain in positions more luminous and/or bluer than the main sequence, except for S1031, which appears on the main sequence in the $K$ vs. $J-K$ diagram. Its position in an optical CMD is likely due to the contribution of its white dwarf companion (see Section 2.2). 

\begin{deluxetable*}{cccccccccccc}[t]
\tablecaption{Basic Stellar Properties of Our Sample of Binaries \label{table:1}}
\tabletypesize
\footnotesize
\tablehead{
\colhead{Sanders} & \colhead{WOCS} & \colhead{Spec.} & \colhead{R.A.} & \colhead{Dec.} & \colhead{$V$\tablenotemark{b}} & \colhead{$(B-V)$\tablenotemark{b}} & \colhead{PRV\tablenotemark{c}} & \colhead{PPM} & \colhead{P} & \colhead{$e$} & \colhead{UV} \\
\multicolumn1c{No.\tablenotemark{a}} & \colhead{ID} & \colhead{Type} & \colhead{(2000)} & \colhead{(2000)} & \colhead{} & \colhead{} & \colhead{} & \colhead{} & \colhead{(days)} & \colhead{} & \colhead{Detection}
}
\startdata
S277 & 2068 & F6 V & 8:49:21.48 & 12:04:22.8 & 12.19 &	0.55 & 84 & 93\tablenotemark{d} & 8567\tablenotemark{f} & 0.859\tablenotemark{f} &  BSS+MSTO \vspace{0.1cm}\\
S997 & 5005 & F5 IV & 8:51:19.90 & 11:47:00.4 &	12.13 & 0.46 & 98 & 99\tablenotemark{e} & 4913\tablenotemark{g} & 0.342\tablenotemark{g} &  BSS+WD \vspace{0.1cm} \\
S1031 & 3001 & F7 V & 8:51:22.96 & 11:49:13.1 &	13.26 & 0.46 & 98 & 99\tablenotemark{e} & 128.14\tablenotemark{f} & 0.04\tablenotemark{f} &  BSS+WD \vspace{0.1cm} \\
S975 & 3010 & F5 V & 8:51:14.36 & 11:45:00.5 &	11.08 & 0.43 & 98 & 90\tablenotemark{d} & 1221\tablenotemark{g} & 0.088\tablenotemark{g} & BSS+WD? \vspace{0.1cm} \\
S1195 & 1025 & F2 V & 8:51:37.69 & 11:37:03.8 & 12.28 & 0.38 & 94 & 99\tablenotemark{e} & 1154\tablenotemark{g} & 0.066\tablenotemark{g} & BSS+WD? \vspace{0.1cm} \\
\enddata
\tablenotetext{a}{Sanders identification number from \citet{sanders77}.}
\tablenotetext{b}{$V$-band magnitude and ($B-V$) index from \citetalias{geller15} and references therein.}
\tablenotetext{c}{Radial velocity membership probability from \citetalias{geller15}.}
\tablenotetext{d}{Proper-motion membership probability from \citet{sanders77}.}
\tablenotetext{e}{Proper-motion membership probability from \citet{girard89}.}
\tablenotetext{f}{Period, eccentricity, and UV detection from \citet{leiner19}.}
\tablenotetext{g}{Period and eccentricity from \citet{latham96}.}
\end{deluxetable*}

\subsection{Descriptions of the Stars to be Analyzed} \label{starDescriptions}
Many of M67's blue stragglers have been analyzed previously; below we review what is known about the five binary systems in our sample:

\hspace{1cm}

\textbf{S277 (WOCS 2068)} is identified as a single-lined binary member \citepalias{geller15} of spectral type F6V \citep{pickles98}. \citetalias{geller15} found S277's radial velocity membership probability to be 84\%, while \citet{sanders77} found S277's proper-motion membership probability to be 93\%. S277 was not included in \citet{gao18}'s RF membership probability study. This star was included in the \citet{deng99} M67 BSS sample of 24 blue stragglers and is a long-period (8567 days), highly eccentric ($e$ = 0.859) member with a binary mass function of 0.0681 \citep{leiner19}. \citetalias{geller15} noted that S277 is located above the turnoff, but lies in an area that is expected to be populated by binaries containing normal main sequence turnoff stars. For this reason, S277 was excluded from the blue straggler sample in \citetalias{geller15}. S277 was included in \citet{leiner19}'s list of blue lurkers in M67, described by \citet{leiner19} to be lower-luminosity counterparts to the BSSs and identified by their anomalous rapid rotation given their age.

\citet{leiner19} found the ultraviolet (UV) flux of S277 was described best by two stars: one starting to turn off the main sequence, and one a blue straggler with a temperature of $\sim$6800 K. They suggested that the system had been through a stellar dynamical encounter that resulted in the collision of two main sequence stars, or a merger in a triple system. \citep{leiner19} found no evidence of a white dwarf in this binary system, and noted that it is unlikely for a white dwarf to have been ejected. \citet{nine23} found an indication of a main sequence stellar companion in the SED of S277 from their \emph{Hubble Space Telescope} (HST) far-UV survey of the M67 blue lurkers. A mass function too large to correspond with a white dwarf companion was reported. The composition of this star has not previously been analyzed.

\hspace{1cm}

\defcitealias{girard89}{Girard et al. 1989}

\textbf{S997 (WOCS 5005)} is a single-lined, long-period binary of 4913 days with high eccentricity ($e$~=~$0.342\pm0.082$) \citep{latham96}. This F5 IV star \citep{allen95} is classified as a blue straggler by both \citetalias{geller15} and \citet{deng99}. The radial velocity membership probability is 98\% and the proper-motion membership probability is 99\% \citepalias{geller15, girard89}. S997 was not included in \citet{gao18}'s RF membership probability study. Formation via mass transfer in a binary system is most likely ruled out due to the high eccentricity of the system \citep{shetrone00, latham96}. The blue straggler is the primary in this system, leaving a binary merger doubtful; this system would have to tidally capture a star after a merger, or be a triple system \citep{shetrone00, leonard96}. 

\citet{vandenberg04} found S997 to be an X-ray source from Chandra observations and speculated that a close binary may be responsible for the X-rays. S997 was also observed to be UV-bright from \emph{Galaxy Evolution Explorer} observations of M67 \citep{sindhu18}. \cite{pandey21} detected a hot companion to this blue straggler using far-UV photometry obtained with the Ultra Violet Imaging Telescope (UVIT on \emph{AstroSat}). Using white dwarf models, they categorized this hot companion to be a white dwarf with $T_{eff} = 13,000 \pm 125$, log($g$)=7.75, $L$/\(L_\odot\) = $0.032 \pm 0.006$, and $R$/\(R_\odot\) = $0.035$$\pm$$0.003$. Additionally, S997 was observed with the Space Telescope Imaging Spectrograph on the HST at low spectral resolution \citep{pandey21}. The Mg II \emph{h+k} lines, which are known as excellent probes of the upper chromosphere \citep[e.g.,][]{leenaarts13}, were inspected in the HST spectrum of S997. These lines were found in absorption, indicating the absence of or low chromospheric activity \citep{pandey21, sindhu18}. The presence of  Mg II \emph{h+k} emission lines would have suggested significant chromospheric activity \citep{sindhu18}. The HST spectra gave additional evidence of FUV excess and confirmation of the presence of both a cool and a hot component in the system \citep{pandey21}. S997 has been previously chemically analyzed by \citet{shetrone00}, who suggested S997 may be a collision product after detecting C and O abundances comparable to turnoff stars.

\hspace{1cm}

\defcitealias{leiner19}{Leiner et al. 2019}

\textbf{S1031 (WOCS 3001)} lies on the blue side of the main sequence turnoff and SIMBAD\footnote{http://simbad.u-strasbg.fr/simbad/} indicates a spectral type of F7V. Despite its blue location, this star was not included in the \citet{deng99} or \citetalias{geller15} candidate BSS lists. \citetalias{geller15} left S1031 out from their sample due to its proximity to the blue hook region on a CMD, so it could not be confidently classified as one of M67's blue stragglers \citepalias{geller15, leiner19}. S1031 was included in \citet{leiner19}'s list of M67 blue lurkers. \citetalias{geller15} classified S1031 as a single-lined binary with a radial velocity membership probability of 98\%. \citet{girard89} estimated its proper-motion membership probability to be 99\%. S1031 was included in \citet{gao18}'s RF cluster membership study, who found the membership probability to be 99.3\%. 

\citet{jadhav19} observed this binary system using the UVIT and suggested the observed excess of UV flux was the result of a white dwarf companion of 12,500 K and $0.3-0.45$ \(M_\odot\). They noted that while the $0.3-0.45$ \(M_\odot\) white dwarf may or may not require mass transfer, the circularised orbit and rapid rotation are indicative of mass transfer in close binaries. \citet{leiner19} found $v$ sin $i$ = 14.7 km s$^{−1}$ , $P$ = 128.14 days, $e$ = 0.04 and a binary mass function = 0.0143. \citet{leiner19} noted that the orbital period of just a few hundred days suggests this system could be the result of Case B mass transfer from a RGB donor. The detection of S1031's high rotation rate gives further evidence of its relation to the blue stragglers \citep{leiner19}. \citet{nine23} reported evidence for a 10,300 – 10,500 K white dwarf companion of 0.36 - 0.43 \(M_\odot\) to S1031 from their HST UV observations. Their best-fit temperatures corresponded to a cooling age of $\sim$600–900 Myr. Based on the orbital properties of the system, \citet{nine23} suggested the companion is a He white dwarf that is the result of Case B mass transfer, though FUV spectroscopy is needed for confirmation. This star has not previously been analyzed for composition.

\hspace{1cm}

\defcitealias{sanders77}{Sanders 1977}

\textbf{S975 (WOCS 3010)} is a single-lined binary star \citepalias{geller15} of spectral type F5 V \citep{pickles98} with rapid rotation ($v$ sin $i$ = 50 km s$^{-1}$) \citep{latham96}. Both \citet{deng99} and \citetalias{geller15} include S975 in their BSS samples. S975's proper-motion membership probability is 90\% and its radial velocity membership probability is 98\% \citepalias{sanders77, geller15}. S975 was not included in \citet{gao18}'s RF membership probability study. 
    
This star was previously chemically analyzed by \citet{shetrone00} who measured a $v$ sin $i$ of 48 km s$^{-1}$. They suggested this was a mass transfer system as their measured O abundance was high in comparison to turnoff stars, but noted the system appeared to have a companion that affected the line flux. S975's orbital solution was calculated by \citet{latham96}, who found the period to be 1221 days and the eccentricity to be small ($e = 0.088$$\pm$$0.060$). The long period and small eccentricity may be an indication of a past interaction. \citet{preston00} suggested the blue stragglers in their blue metal-poor star sample were formed by mass transfer, as they had long periods and small orbital eccentricities characteristic of the McClure carbon star binaries (see the bottom panel of Figure 19 in \citet{preston00}). The carbon star binaries of McClure \citep{mcclure1997, mcclure90} exhibited long periods and low orbital eccentricities, which McClure explained by orbit dissipation occurring with mass transfer in the system. \citet{leonard96} suggested the formation of S975 may be due to Case C (asymptotic giant branch) mass transfer. \citet{landsman98} reports strong evidence of a hot subluminous companion to S975 in their UV photometry. \citet{landsman98} suggests that the existence of a hot subluminous companion along with a nearly circular binary orbit indicates that the formation is due to a binary mass-transfer process. \citep{sindhu19} detected a UV excess consistent with this sub-luminous, hot companion through SED fitting of S975. The hot component has a $T_{eff}$ = $13,750 - 16,500$ K, $R$/\(R_\odot\) = $0.047 - 0.029$, and mass $\sim$0.18\(M_\odot\), suggesting the companion is a white dwarf.

\hspace{1cm}

\textbf{S1195 (WOCS 1025)} is defined as a single-lined binary star \citepalias{geller15} of spectral type F2 V \citep{pickles98}. This star was one of the 24 candidate BSSs from \citet{deng99} and included in the \citetalias{geller15} blue straggler candidate sample. \citet{girard89} estimated its proper-motion membership probability to be 99\% while \citetalias{geller15} estimated its radial velocity membership to be 94\%. \citet{gao18} found S1195's RF membership probability to be 99\%. \citet{latham96} estimated the period to be 1154 days, the eccentricity to be 0.066$\pm$0.082, and the $v$ sin $i$ to be 60 km s$^{-1}$. The circularised orbit paired with a large $v$ sin $i$ suggests recent mass transfer episodes for this single-lined binary \citep{pandey21}. 

When observing this system, \citet{pandey21} detected a UV excess and reported that this blue straggler likely has a hot companion with a $T_{eff}$ in the range $21,000 - 50,000$ K, but more data in the UV are needed to confidently estimate its parameters. This star has not previously been analyzed for composition. 

\subsection{IGRINS Spectra}

Spectroscopic observations of the five stars were carried out with the high-resolution $H$ and $K$ photometric band Immersion Grating Infrared Spectrograph \citep[IGRINS,][]{mace16,park14,yuk10}\footnote{http://www.as.utexas.edu/astronomy/research/people/jaffe/    igrins.html} at the McDonald Observatory 2.7m Harlan J. Smith Telescope. Target acquisition was straightforward and there was no source confusion in the field of any target. 

A single IGRINS observation delivers full spectral coverage of the $H$ ($\sim$1.5$-$1.7~$\mu$m) and $K$ ($\sim$2.0$-$2.4~$\mu$m) bands with resolving power $R$~$\equiv$ $\lambda/\Delta\lambda$~$\simeq$ 45,000. Each BSS target observation was accompanied by observation of a hot, mostly featureless star usually of spectral type A0V to be used in the removal of telluric lines.

Each complete observation of a target star consisted of 1$-$2 sets of four individual integrations taken in an A-B-B-A pattern, where A and B integrations were accomplished with the target shifted on the slit perpendicular to the dispersion direction by 7~arcseconds. The exposure time for each integration was 600~seconds, resulting in total exposure times of 4800~seconds (80~minutes) for all BSS stars except S975, for which only one set of four integrations was accomplished. Exposure times for the hot-star observations varied but were always adequate to attain good signal-to-noise ($S/N$) in the reduced spectra.

\begin{figure}[t]
\epsscale{1.178}
\plotone{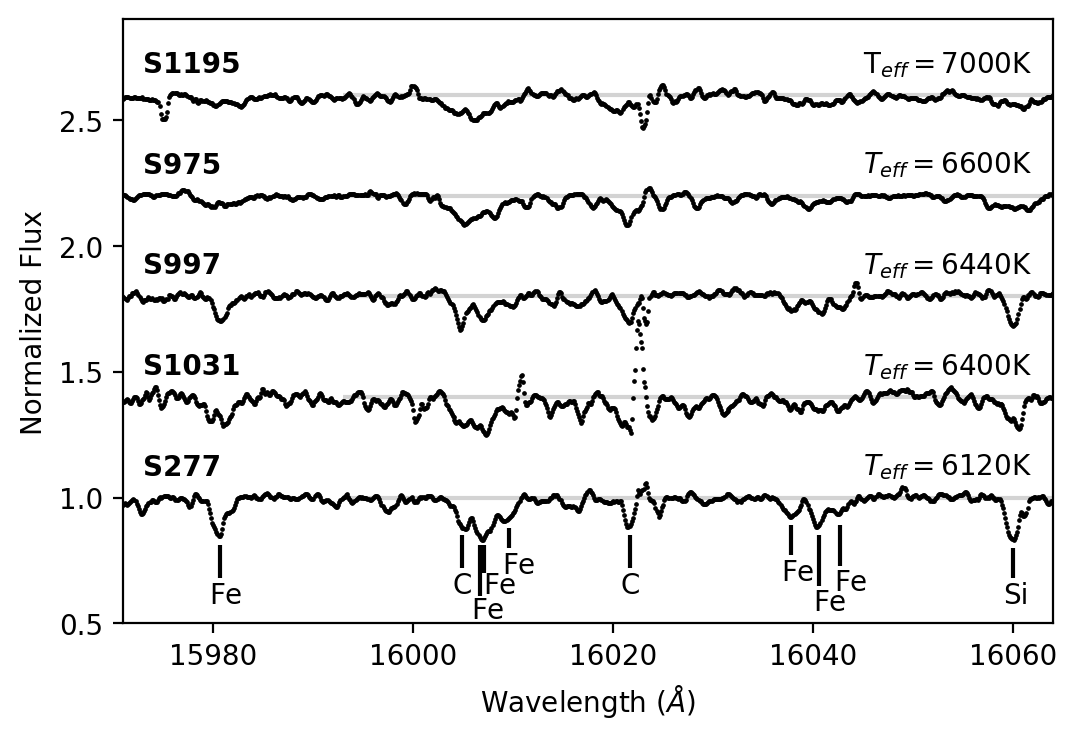}
\caption{A sample spectral region in the $H$ band. The spectra are arranged in order of decreasing $T_{eff}$ from top to bottom. The rotational broadening of the spectral lines can be seen in S1031, S975, and S1195. Derived rotational velocities are shown in Table \ref{table:3}. \label{figure:3}}
\end{figure}

The IGRINS Pipeline Package\footnote{https://github.com/igrins/plp} was used to reduce the data \citep{lee16}. The pipeline performs flat-fielding, background removal, order extraction, distortion correction, wavelength calibration, and telluric removal. To normalize the data, each spectrum was divided by a cubic spline fit to the continuum. Figure \ref{figure:3} shows a spectral region from $15970-16065$ \AA \space for our five targets.

\section{ANALYSIS AND RESULTS}\label{modabund}

In this section we detail our line list construction, calculation of atmospheric parameters, and abundance measurement methods. The results of our abundance analysis are presented and compared to results from previous chemical studies. The reliability of the measured spectral lines is evaluated in terms of departures from local thermodynamic equilibrium (LTE).

\subsection{Line List}
Atomic line data for the spectral lines come from \citet{afsar18} and the atomic and molecular line list generator \textit{linemake} \citep{placco21}. The log($gf$) values were calibrated two ways, first by fitting a high-resolution solar spectrum with a synthetic spectrum and, second, by adjusting MOOG's equivalent width line analysis \citep[][V. 2017]{sneden73} to produce solar abundances. Solar abundances were adopted from \cite{2009ARA&A..47..481A}. The averages of the resulting log($gf$) values from the two procedures were taken to be the final log($gf$) values for the lines. Table \ref{table:2} shows the final line list adopted. 

We note that the lines of C I arise from high excitation levels and are likely formed in non-LTE (NLTE) \citep{fab06}. By using log($gf$) values derived from fitting these lines in the solar spectrum to obtain the known solar [C/H] abundance with a LTE model, we are implicitly including an approximate NLTE correction at the solar effective temperature. 

\begin{deluxetable}{ccccc}[t]
\tablecaption{Line List\label{table:2}}
\tablehead{
\colhead{Element} & \colhead{$\lambda_{vacuum}$ (\AA)} & \colhead{$\chi$ (eV)} & \colhead{log $gf$} & \colhead{Source\tablenotemark{a}}
}
\startdata
C I & 16004.9 & 9.624 & 0.18 & 2 \vspace{0.1cm}\\
C I & 16021.7 & 9.63 & 0.15 & 1 \vspace{0.1cm} \\
C I & 16890.386 & 9 & 0.355 & 1 \vspace{0.1cm} \\
C I & 17234.473 & 9.69 & 0.113 & 2 \vspace{0.1cm} \\
C I & 17274.945 & 9.688 & 0.325 & 2 \vspace{0.1cm}\\
C I & 21023.131 & 9.17 & -0.44 & 1 \vspace{0.1cm} \\
C I & 21191.345 & 9.827 & -0.25 & 2 \vspace{0.1cm} \\
C I & 21191.839 & 9.827 & -0.48 & 2 \vspace{0.1cm} \\
C I & 21259.256 & 9.826 & -0.8 & 2 \vspace{0.1cm} \\
C I & 21259.934 & 9.826 & 0.3 & 2 \vspace{0.1cm} \\
... & & & &
\enddata
\tablenotetext{a}{Sources for the excitation potential values: 1. \citet{afsar18}; 2. \citet{placco21}.}
\tablecomments{This table is available in its entirety in a machine-readable form in the online journal.}
\end{deluxetable}

\subsection{Effective Temperatures}
We calculated the photometric temperature of the stars using the $T_{eff}$ and color relation from \citet{2009A&A...497..497G}: 
\begin{multline}
    \theta_{eff} = b_0 + b_1 X + b_2 X^2 + b_3 X[Fe/H] + b_4[Fe/H] \\ + b_5[Fe/H]^2,
\end{multline}
where $\theta_{eff} = 5040/T_{eff}$, $X$ is the color, and $b_i$ ($i=0,...,5$) are the coefficients of the fit. The effective temperature for each star was calculated four times using $E(B-V)$ from \citep{bellini10} and the stars' $B-V$, $V-J$, $V-H$, and $V-K$ colors. The $b_i$ coefficients for dwarf stars were used. The final effective temperatures are the averages of the four photometric temperatures for each star. These effective temperatures and their standard deviations are shown in Table \ref{table:3}. Two of our program stars, S975 and S1195, show much larger differences between the $B-V$ temperature and the $V-J$, $V-H$, and $V-K$ temperatures. The differences between the $B-V$ temperature and the average of the $V-J$, $V-H$, and $V-K$ temperatures were $\sim$ 1040 K and $\sim$ 680 K, respectively. Due to this dispersion in the effective temperatures from the four colors, we used an average of the effective temperatures from \citet{liu08} and references therein for S975 and S1195. The effective temperatures from \citet{liu08} are consistent with the calculated $V-J$, $V-H$, and $V-K$ temperatures. \citet{liu08} also gives effective temperatures for S277 and S997, which are consistent with the effective temperatures adopted in our study.

\subsection{Surface Gravities}

The standard equation for physical gravities depends on mass:
\begin{multline}
  \logg_{\star} = 0.4 (M_{\rm V\star} + BC - M_{\rm Bol\sun}) + \logg_{\sun} \\ + 4 {\rm log} (\frac{\teff_{\star}}{\teff_{\sun}}) + {\rm log} (\frac{{\it m}_{\star}}{{\it m}_{\sun}}).
\end{multline}
Previous studies of BSSs in open clusters have found them to be a few tenths of a solar mass greater than the clusters’ turnoff mass.  For example, \citet{jadhav21} found BSS masses between $1.2-1.9$ $M_{\rm \sun}$ in the old open cluster King 2.  \citet{sandquist21} determined a turnoff mass of 1.22 $M_{\rm \sun}$ for M67, and a mass of 1.32 $M_{\rm \sun}$ for the bluest subgiants in the cluster. \citet{stello16} found masses of 1.36 $M_{\rm \sun}$ for M67 giants from asteroseismology. 

Since the masses of our specific targets are uncertain, we adopted surface gravities from \citet{liu08} for S975 and S1195. When the surface gravities for S975 and S1195 are computed using Equation 2, log $g$ is equal to 3.13 and 3.85, respectively. S1031 lies near the main sequence, slightly below the turnoff, and we adopted a mass of 1.1 $M_{\rm \sun}$ for a surface gravity of log $g$ = 4.1.  S997 and S277 both lie about half a magnitude above the early subgiant sequence, and redward of the extension of the main sequence. For these stars we adopted surface gravities log $g$ = 3.6, consistent with possible mass ranges and their evolutionary state. We note that the sensitivity of our derived abundances to a change in surface gravity of $\Delta$log $g$ = +0.3 is typically -0.01 or -0.02 dex in Fe in abundance. Neither the adopted stellar mass nor the surface gravity has a significant effect on our results. The adopted surface gravity for each star is included in Table \ref{table:3}.

\begin{deluxetable}{cccccc}[t]
\tablecaption{Derived Stellar Properties\label{table:3}}
\tablehead{
\colhead{Sanders} & \colhead{WOCS} &  \colhead{$T_{eff}$} & \colhead{log $g$} & \colhead{$\xi$} & \colhead{$v$ sin $i$} \\
\multicolumn1c{No.} & \colhead{ID} & \colhead{(K)} & \colhead{} & \colhead{(km s$^{-1}$)} &\colhead{(km s$^{-1}$)} 
}
\startdata
S277 & 2068 & 6120$\pm$121 & 3.6 & 2.7 & $<$14 \vspace{0.1cm}\\
S997 & 5005 & 6440$\pm$48 & 3.6 & 1.8 & $<$20 \vspace{0.1cm} \\
S1031 & 3001 & 6400$\pm$34 & 4.1 & 1.3 & 25$\pm$6 \vspace{0.1cm} \\
S975 & 3010 & 6600\tablenotemark{a} & 4.1\tablenotemark{a} & 1.6 & 51$\pm$5 \vspace{0.1cm} \\
S1195 & 1025 & 7000\tablenotemark{a} & 4.5\tablenotemark{a} & 2.3 & 58$\pm$6 \vspace{0.1cm} \\
\enddata
\tablenotetext{a}{Values from \citet{liu08} and references therein.}
\end{deluxetable}

\subsection{Microturbulent Velocities}
For two of the stars in our sample, S277 and S997, the spectra allowed for detailed equivalent width analysis as the line profiles were not rotationally broadened. For these stars, the microturbulence was derived from abundance vs. line strength plots for the measured Fe lines, using the adopted effective temperature and surface gravity. For the remaining stars S1031, S975, and S1195 with spectra that were more blended, microturbulent velocities were calculated using the relation from \citet{gebran14}:
\begin{equation}
    \xi_t = 3.31 \times exp[-(log(\frac{T_{eff}}{8071.03})^2/0.01045)], 
\end{equation}
which applies to our sample of F-type stars. This relation gives microturbulent velocities of 0.83 km s$^{-1}$ for S277 and 1.3 km s$^{-1}$ for S997. Microturbulent velocities do not dominate the uncertainty of the derived abundances in the BSSs, so the differences in derived microturbulent velocities from the two approaches do not have a significant impact. Final microturbulent velocities are shown in Table \ref{table:3}. 

\subsection{Assumed Metallicities}
M67 has been studied at high spectral resolution many times and all studies agree that it has essentially solar metallicity. A sample of published values in recent years includes $-$0.03~$\pm$~0.03 \citep{tautvaisiene00}, $+$0.02~$\pm$~0.04 \citep{yong05}, $+$0.03~$\pm$~0.03 \citep{randich06}, $+$0.01~$\pm$~0.03 \citep{santos09}, $+$0.05~$\pm$~0.02 \citep{pancino10}, $-$0.01~$\pm$~0.05 \citep{jacobson11}, $+$0.04~$\pm$~0.02 \citep{hawkins16}, and $+$0.03~$\pm$~0.05 \citep{netopil16}. Therefore, we adopt [Fe/H]~=~+0.02 for this study, the average of published values.

\begin{deluxetable*}{ccccccccccccccccccccccc}[ht]
\tablecaption{Stellar Abundances\label{table:4}}
\tablehead{
\colhead{} & \colhead{} & \colhead{}  & \colhead{S277} & \colhead{} & \colhead{}  & \colhead{} & \colhead{S997} & \colhead{} & \colhead{}  & \colhead{} & \colhead{S1031} & \colhead{} & \colhead{}  & \colhead{} & \colhead{S975} & \colhead{} & \colhead{}  & \colhead{} & \colhead{S1195} & \colhead{} & \colhead{} & \colhead{} \\
\cline{3-5}
\cline{7-9}
\cline{11-13}
\cline{15-17}
\cline{19-21}
\multicolumn2c{Element}  & \colhead{[M/H]} & \colhead{$\sigma$} & \colhead{n} & \colhead{} &\colhead{[M/H]} & \colhead{$\sigma$} & \colhead{n} & \colhead{} &\colhead{[M/H]} & \colhead{$\sigma$} & \colhead{n} & \colhead{} &\colhead{[M/H]} & \colhead{$\sigma$} & \colhead{n} & \colhead{} &\colhead{[M/H]} & \colhead{$\sigma$} & \colhead{n} & \colhead{Avg.} & \colhead{$\sigma$} 
}
\startdata
C\tablenotemark{a} & & -0.22 & 0.08 & 10 & & -0.09 & 0.11 & 5 & & 0.11 & 0.26 & 3 & & 0.24 & 0.19 & 3 & & 0.17 & 0.27 & 5 & 0.04 & 0.20 \vspace{0.1cm} \\
Na & & 0.24 & 0.14 & 1 & & 0.00 & 0.17 & 1 & & - & - & - & & - & - & - & & - & - & - & 0.12 & 0.16 \vspace{0.1cm} \\
Mg & & 0.05 & 0.12 & 1 & & 0.30 & 0.12 & 2 & & 0.08 & 0.26 & 2 & & -0.10 & 0.19 & 2 & & -0.10 & 0.23 & 2 & 0.05 & 0.19 \vspace{0.1cm} \\
Al & & 0.24 & 0.14 & 3 & & -0.09 & 0.13 & 2 & & 0.17 & 0.23 & 2 & & -0.03 &  0.19 & 2 & & - & - & - & 0.07 & 0.18 \vspace{0.1cm} \\
Si & & 0.06 & 0.16 & 11 & & 0.22 & 0.11 & 9 & & 0.35 & 0.23 & 8 & & 0.13 & 0.18 & 3 & & 0.00 & 0.24 & 3 & 0.15 & 0.19 \vspace{0.1cm} \\
S & & 0.15 & 0.13 & 2 & & 0.03 & 0.14 & 2 & & 0.05 & 0.28 & 1 & & 0.27 & 0.20 & 2 & & - & - & - & 0.13 & 0.20 \vspace{0.1cm} \\
Ca & & 0.10 & 0.09 & 7 & & 0.31 & 0.14 & 6 & & -0.02 & 0.25 & 3 & & 0.00 & 0.18 & 2 & & - & - & - & 0.10 & 0.18 \vspace{0.1cm} \\
Fe & & 0.00 & 0.06 & 46 & & -0.01 & 0.10 & 46 & & 0.01 & 0.23 & 23 & & -0.10 &  0.19 & 30 & & 0.03 & 0.23 & 24 & -0.01 & 0.18 \vspace{0.1cm} \\
Ni & & -0.16 & 0.12 & 2 & & -0.10 & 0.18 & 2 & & 0.30 & 0.27 & 2 & & - & - & - & & - & - & - & 0.01& 0.20 \vspace{0.1cm} \\
\enddata
\tablenotetext{a}{Differentially corrected for NLTE using \citet{fab06} results.}
\tablecomments{ n = number of lines}
\end{deluxetable*}

\subsection{Abundance Fitting and Results}

The analysis method for each spectral line was determined individually by the degree of blending of each profile. For blended lines, spectra were synthesized using the current version of the local thermodynamic equilibrium, plane parallel, spectral line analysis code MOOG along with MARCS 1-D atmospheric models \citep{gustafsson08}. Synthetic spectra were generated using the derived atmospheric parameters and overplotted on the observed spectra. Fits were visually adjusted. For lines that were not blended, the equivalent widths were measured with the IRAF task \textit{splot}. Abundances of each transition were derived by requiring computed and measured equivalent widths to agree. The final elemental abundances of C, Na, Mg, Al, Si, S, Ca, Fe, and Ni are shown in Table \ref{table:4}. Column 1 is the observed element. Subsequent columns show the abundance [M/H] for each element. Solar log $\epsilon$(A)\footnote{We use the standard spectroscopic notation where log $\epsilon$(A) $\equiv$ log($N_A$/$N_H$)+12.0 and [$A/B$] $\equiv$ log($N_A$/$N_B$)$_{star}$ $-$ log($N_A$/$N_B$)$_{\odot}$ for elements $A$ and $B$.} values from \citet{2009ARA&A..47..481A} were used to determine [M/H] values. The standard deviation along with the number of lines are also given. Final uncertainties were computed from the sum in quadrature of the individual uncertainties for the atmospheric parameters and observational uncertainty. Average surface abundances and standard deviations for the set of program stars are shown in the last two columns. Due to line weakening and/or smearing from the high temperature or rapid rotation of S1031, S975, and S1195, some of the elemental abundances could not be measured.

\begin{figure*}[h]
\epsscale{1.174}
\plotone{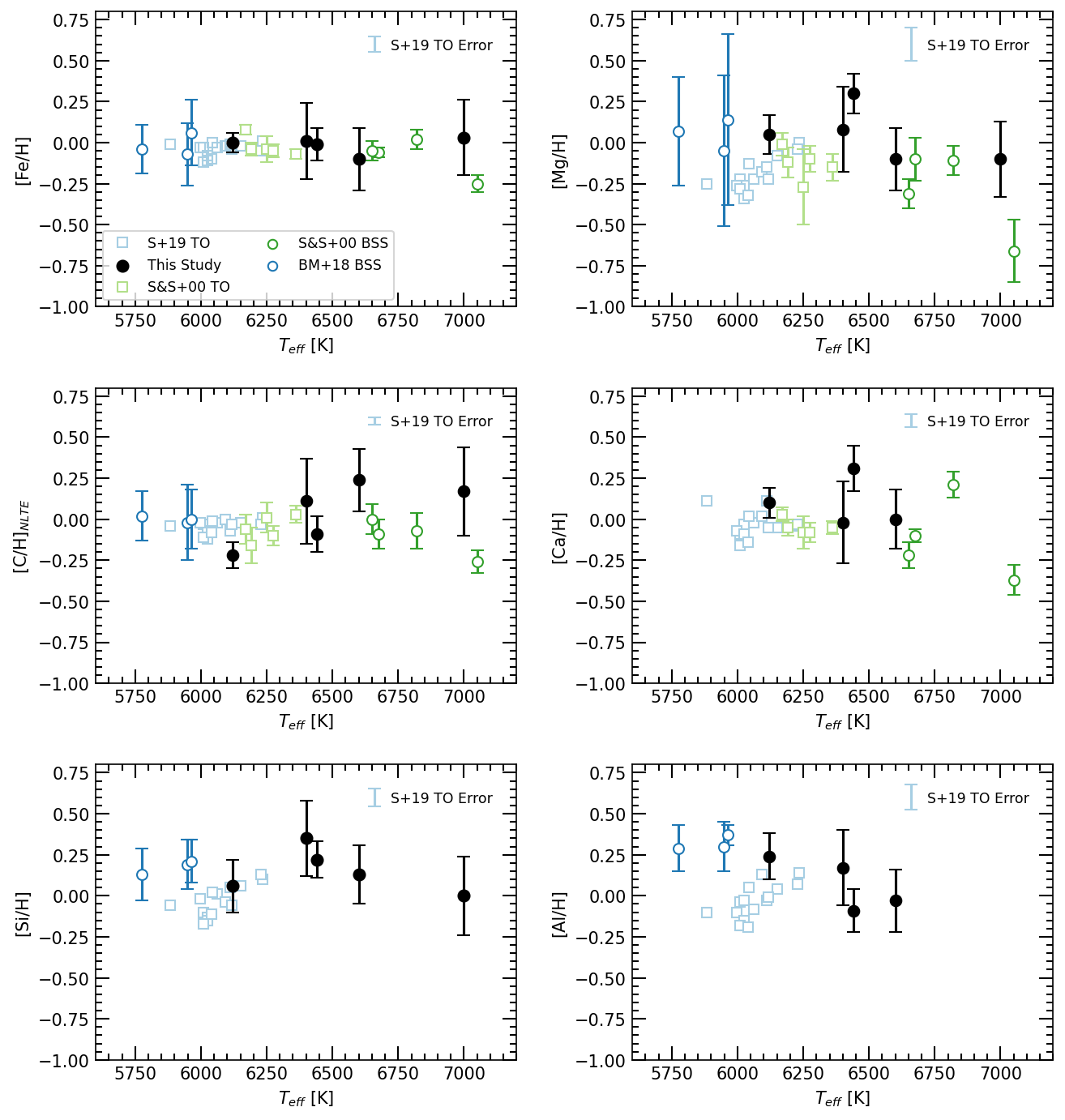}
\caption{Chemical abundance vs. effective temperature. The blue straggler stars from our analysis are shown as black circles. BSS error bars are one standard deviation. M67 turnoff star abundances from \citet{souto19} are shown as blue squares and from \citet{shetrone00} as green squares. Uncertainties from \citet{souto19} are shown in the upper right corner of each panel. M67 BSS abundances from \citet{motta18} are blue circles and from \citet{shetrone00} are green circles. Data from the literature are unfilled symbols. \label{figure:4}}
\end{figure*}

The surface abundances of the program stars are generally consistent with those of turnoff stars. Figure \ref{figure:4} shows the derived abundances for six of our elements as a function of effective temperature. Abundance measurements of turnoff stars in M67 from \citet{shetrone00} and from APOGEE\footnote{Apache Point Observatory Galactic Evolution Experiment: https://www.sdss.org/dr16/irspec/} spectra from \citet{souto19} are plotted for comparison. In particular, our iron abundance does not significantly differ from the cluster average. Our results for Al and Ca are consistent with the average values for turnoff stars. Measured Si and Mg abundances are slightly higher than turnoff star abundances, but are within the errors. Carbon abundances are discussed further in Section 3.7. APOGEE results for Mg, Si, and Al all suggest a strong temperature dependence, but our results do not.

Our results are similar to the previous M67 blue straggler optical chemical composition studies of \citet{mathys91}, \citet{shetrone00}, and \citet{motta18}, who also found the surface abundances of their blue straggler stars to be in agreement with turnoff star abundances. \citet{shetrone00} and \citet{motta18} BSS abundances are shown in Figure \ref{figure:4} for comparison. 

Two stars in our sample, S997 and S975, were previously analyzed by \citet{shetrone00}. Their analysis was based on spectra covering from 3800\AA \space to 10100\AA. \citet{shetrone00} photometrically determined their effective temperatures using the equation from \citet{soderblom93}, then fine tuned the temperatures by forcing the slope of abundances from Fe I lines vs. excitation potential to be zero. Their final effective temperatures are $\sim$200 K greater than our final effective temperatures. For S997 and S975, the sensitivity of our derived abundances to a change in temperature of $\pm$100 K is less than 0.1 dex
in abundance. The surface gravities and microturbulances for both S997 and S975 in \citet{shetrone00} are also higher than our values. C, Mg, Ni, Na, and Ca abundances were measured in both of our studies. Within the precision of both studies, our surface abundance measurements agree with those of \citet{shetrone00}, with the exception of [C/H] in S975 and [Mg/H] and [Ca/H] in S997, which we find to be higher. Differences in derived carbon abundances are discussed in Section 3.7.

\subsection{C I Line Formation}
The  C I lines used to determine the abundance of carbon in the BSSs are all high-excitation lines with excitation potentials of 9 eV or above. Typical equivalent widths of the carbon lines are around log($\omega/\lambda$) = $-4.9$. \citet{amarsi19} investigated 3D NLTE departures from simple, LTE line formation for high-excitation C I lines in the solar photosphere, and found that corrections for near-IR lines are typically 0.05 dex or less for the disk-integrated flux. While the NLTE corrections can be modest, neglecting neutral hydrogen impact excitation can create larger deviations from a LTE model \citep{amarsi19}. \citet{tomkin92} included a NLTE analysis of the the 9100 \AA \space C I lines used in their abundance study of halo dwarfs. They found NLTE corrections for the 7.48 eV C I lines to be on the order of 0.05 dex, but observed an effective temperature dependence in the C abundances. They noted the temperature dependence may be due to the representation of the NLTE effects, and specifically, the treatment of neutral hydrogen collisions. \citet{fab06} provided neutral carbon NLTE corrections for the $\sim$ 7.5 eV lines at approximately 9000 \AA, and reported large, negative NLTE abundance corrections for high-excitation neutral carbon permitted lines. 

As a simple check on the reliability of the C I lines used in our analysis, we selected four main sequence stars from the IGRINS spectral library \citep{park18} covering a range of temperature from 5801 $\leq$ $T_{eff}$ $\leq$ 6573 K. The spectra, observed with the same instrument and resolution as our BSSs, have S/N ratios of typically S/N $\geq$ 200. We derived [C/H] abundances using the atmospheric parameters adopted by \citet{park18} to identify any strong dependence of [C/Fe] on temperature that might indicate a departure from LTE. No dependence of [C/Fe] on temperature was found, though all of the stars are unexpectedly C-rich. The derived [C/H] abundances are given in Table \ref{table:5}.

\begin{deluxetable}{cccccc}[t]
\tablecaption{IGRINS Spectral Library Star Abundances\label{table:5}}
\tablehead{
\colhead{Star} & \colhead{Spectral} & \colhead{$T_{eff}$\tablenotemark{a}} & \colhead{[Fe/H]\tablenotemark{a}} & \colhead{log g\tablenotemark{a}} & \colhead{[C/H]} \\
\multicolumn1c{} & \colhead{Type} & \colhead{(K)} & \colhead{} & \colhead{} &\colhead{} 
}
\startdata
HD71148 & G1 V & 5801 & -0.04 & 4.36 & 0.31$\pm$0.14 \vspace{0.1cm}\\
HD87141 & F5 V & 6359 & 0.09 & 3.90 & 0.18$\pm$0.14 \vspace{0.1cm} \\
HD91752 & F3 V & 6418 & -0.23 & 3.96 & 0.04$\pm$0.14 \vspace{0.1cm} \\
HD87822 & F4 V & 6573 & 0.10 & 4.06 & 0.35$\pm$0.14  \vspace{0.1cm} \\
\enddata
\tablenotetext{a}{Stellar atmospheric parameters are from \citet{park18} and references therein.}
\end{deluxetable}

NLTE corrections were not applied to the $\sim$ 8.6 eV carbon lines in \citet{shetrone00}, and it is unclear what corrections might be appropriate for these carbon lines or the 9+ eV lines used in our study. In \citet{ferraro06}'s chemical analysis of BSSs in the globular cluster 47 Tuc, a NLTE correction was derived by interpolating the C I abundances from \citet{tomkin92} and applied to the 7.48 eV C I line at 9111.8 \AA. The large corrections for the 7.5 eV lines from \citet{fab06} suggest significant NLTE corrections are likely, on the order of -0.2 to -0.4 dex depending on the effective temperatures and surface gravities of our program stars.

We estimate neutral carbon NLTE corrections for our program stars with the \citet{fab06} results to provide differential corrections compared to the Sun. The log($gf$) values in our line list were derived from comparison with the solar spectrum. Therefore, spectral analysis with these log($gf$) values includes the solar correction for NLTE, resulting in an abundance that is partially corrected for NLTE. Differential corrections are then made with the \citet{fab06} results based on the temperature and gravity of the Sun and our program stars. When the \citet{fab06} corrections are applied to our BSS and IGRINS spectral library star carbon abundance measurements, the average carbon abundance in both groups is reduced by $\sim$ 0.1 dex. The final BSS carbon abundances that are differentially corrected for NLTE are shown in Table \ref{table:4}.

In comparison to our NLTE corrected carbon abundances, the carbon abundance for S997 derived by \citet{shetrone00} is the same, while for S975 our result is higher by $\sim$ 0.3 dex. We adopted \citet{shetrone00}'s atmospheric parameters and repeated our carbon abundance analysis of our program stars. For S997, we find [C/H]~=~0.02$\pm$0.11. This is slightly closer to the turnoff value for carbon compared to our [C/H]~=~0.08 result where NLTE corrections from \citet{fab06} were not yet applied. Adopting the \citet{shetrone00} model atmosphere parameters gave the same carbon abundance for S975: [C/H]~=~0.32$\pm$0.19, the carbon abundance before differential NLTE corrections were applied. 

Overall, the C I lines used in this study are new to abundance determinations, are sufficient in strength, and reside in spectral regions free from telluric lines. However, they may be sensitive to 3D and/or NLTE effects, which would result in too large of LTE abundances. We estimate neutral carbon NLTE corrections for our program stars, but a NLTE analysis of these high-excitation lines is needed to better understand their behavior.

\section{DISCUSSION}\label{discuss}

\subsection{Rotation Velocity}

We measure rotational velocities ($v$ sin $i$) for three of our program stars: S975, S1031, and S1195. By fitting the rotational broadening of the spectral lines, we found $v$ sin $i$ $=$ 25$\pm$6 km s$^{-1}$ for S1031, $v$ sin $i$ $=$ 51$\pm$5 km s$^{-1}$ for S975, and $v$ sin $i$ $=$ 58$\pm$6 km s$^{-1}$ for S1195. The rotation detected is comparable to the $v$ sin $i$ values from \citet{nine23} and \citet{leiner19} for S1031 (24.7 km s$^{-1}$and 14.7 km s$^{-1}$, respectively), \citet{shetrone00} for S975 (50 km s$^{-1}$), and \citet{latham96} for S975 (50 km s$^{-1}$) and S1195 (60 km s$^{-1}$).

Rotation rates may evolve with time in mass-transfer objects. Spin-up of the mass-accreting star is expected from significant angular momentum transport during mass transfer \citep{packet81, deMink13, matrozis17}. Likewise, stellar collisions and mergers can result in stars with rapid rotation \citep{sills01,sills05}. \citet{leiner18} recently conducted the first observational study of spin-down in post-mass-transfer binaries, where the stars spin down as they age in a similar form that is predicted for standard solar-type stars. The gyro-age clock is reset and old stars can be seen with rapid rotation typical of younger stars \citep{leiner18}. Rapid rotation rates may be indicative of a recent stellar interaction \citep[e.g.,][]{leiner19}, which is relevant for three of our stars (S975, S1031, and S1195). Unfortunately, the $v$ sin $i$ does not appear to give information on what formation mechanism occurred, as multiple interactions can lead to rapid rotation.

\citet{subramaniam20} found a moderate correlation between the temperature of a white dwarf companion and the $v$ sin $i$ of the blue straggler (see Figure 3 in \citet{subramaniam20}). They concluded that BSSs with faster rotation have hot white dwarf companions, while slower rotating BSSs do not have hot companions. Together with the predicted temperatures of the white dwarf companions, our $v$ sin $i$ measurements support \citet{subramaniam20}’s conclusion that the white dwarf companions of fast-rotating blue stragglers and blue lurkers are hot.

\subsection{Possible Blue Straggler Formation Histories}
The derived surface abundances of the program stars are generally consistent with those of turnoff stars. This is in agreement with the previous M67 BSS abundance analyses of \citet{mathys91}, \citet{shetrone00}, and \citet{motta18}.

Each star in our sample could have a different history, and through the possible detection of a change in initial abundances, carbon has been considered to be an important indicator of what those histories might be. Given the complexity of carbon, we are unable to determine if carbon is enhanced or depleted with certainty, and a [C/Fe] similar to that of turnoff stars is most likely. We detect possible carbon enhancements for S975 and S1195 consistent with predicted AGB-phase mass transfers, possible carbon depletion in S277 inconsistent with the predicted stellar dynamical encounter or a merger in a triple system formation, no carbon enhancement or depletion in S997 which is inconsistent with the predicted mass transfer formation mechanism, and no evidence of depletion of carbon in S1031 from the predicted RGB-phase mass transfer. Ultimately, we are hesitant to use carbon abundances as an indicator of mass transfer due to the large uncertainties associated with NLTE effects. With the inconclusive C abundances, determining the histories of the systems mainly relies on dynamical and white dwarf companion information. The expected range of carbon abundances from different formation scenarios is too small given the observational errors \citep{char10}.

Comparing the location of the BSSs on the IR CMD versus the optical CMD may give insight into the formation mechanisms of the BSSs through detection of the companion's influence on the observed colors. The location of S277 on the infrared CMD is comparable to its optical location, providing evidence in favor of a dynamical collision or merger in a triple system, as a white dwarf that would imply mass transfer is not detected. While S1031 lies blueward of the main sequence turnoff in the optical, it appears on the main sequence in the infrared, suggesting a hot companion is contributing to the star's optical color. S975 remains in a position bluer and more luminous than the main sequence turnoff in the IR CMD, which can be explained by the subluminous nature of the white dwarf companion. The location of S997 on the infrared CMD is comparable to its optical location, confirming the hot companion is faint in nature. No evidence of a hot companion to S1195 was found from the IR CMD. The locations of the BSSs on the IR CMD versus the optical CMD generally support the conclusions drawn from the SED analyses of the individual BSSs described in Section \ref{starDescriptions}.

\section{SUMMARY AND CONCLUSIONS}\label{conclusions}

The first detailed infrared chemical analysis of five binary members of the old open cluster M67 located above and/or blueward of the cluster’s main sequence turnoff is reported. We measured C, Na, Mg, Al, Si, S, Ca, Fe, and Ni abundances in the program stars using high-resolution, infrared spectra covering the $H$ and $K$ bands. Our principal conclusions are as follows:

\begin{enumerate}
    \item The detected compositions of the five anomalous stars in this study are generally consistent with those of turnoff stars in M67. 
    \item Abundances from our program stars determined from infrared spectroscopy generally agree with results obtained from optical spectroscopy. 
    \item Projected rotational velocities determined from infrared spectra generally agree with those from optical spectra.
    \item S1031 is displaced blueward from the main sequence in the $RP$ vs. $(BP-RP)$ but lies on the main sequence in the $K$ vs. $(J-K)$ diagram. Flux from a hot companion may be the cause of the displacement in the $(BP-RP)$ color.
    \item Rapid rotation is detected in S1031, S975, and S1195, which together with the possible carbon enhancements of S975 and S1195 from the IR spectra and location of S1031 on an IR CMD, may be indicative of recent stellar mass transfer in a binary system. 
    \item The near-IR high-excitation neutral carbon lines of 9+ eV are used for the first time in stellar abundance determinations, and require further NLTE analysis to better understand their behavior.
\end{enumerate}

Our spectroscopic analysis of these five binary members does not provide conclusive evidence of altered composition that would be indicative of certain formation mechanisms. The relative decreases of C on the red giant branch of mass transfer that occurs at that phase, are small and difficult to detect. Detection of the enhancements from post-AGB evolution will also be minimal and difficult to see due to the uncertainty in the C abundances. {We are skeptical to use carbon abundances as an indicator of mass transfer due to the large uncertainties associated with NLTE effects and the relatively modest changes that may occur through red giant and asymptomatic giant branch evolution. Derived surface abundances are generally consistent with those of turnoff stars, and elemental abundances that vary significantly from the turnoff values need additional line measurements for definitive conclusion. These results agree with previous optical studies of M67 BSS chemical compositions. 

\section{Acknowledgements }\label{acknowledgements}

This work used the Immersion Grating Infrared Spectrograph (IGRINS) that was developed under a collaboration between the University of Texas at Austin and the Korea Astronomy and Space Science Institute (KASI) with the financial support of the US National Science Foundation under grant AST-1229522, of the University of Texas at Austin, and of the Korean GMT Project of KASI. This research has made use of NASA's Astrophysics Data System Bibliographic Services, the HITRAN database operated by the Center for Astrophysics, and the SIMBAD database, operated at CDS, Strasbourg, France. This publication also makes use of data products from the Two Micron All Sky Survey, which is a joint project of the University of Massachusetts and the Infrared Processing and Analysis Center/California Institute of Technology, funded by the National Aeronautics and Space Administration and the National Science Foundation. Additionally, this work presents results from the European Space Agency (ESA) space mission Gaia. Gaia data are being processed by the Gaia Data Processing and Analysis Consortium (DPAC). Funding for the DPAC is provided by national institutions, in particular the institutions participating in the Gaia MultiLateral Agreement (MLA). Support from the Daniel Kirkwood Endowment at Indiana University is gratefully acknowledged. Finally, we thank the referee for their thoughtful comments that have materially improved our paper.

\facilities{McDonald Observatory 2.7m Harlan J. Smith Telescope}
\software{IRAF \citep{tody86, tody93}, MOOG \citep[][V. 2017]{sneden73}, matplotlib \citep{hunter07}, numpy \citep{vanderwalt11}}

\newpage
\bibliographystyle{aasjournal}



\end{document}